\documentclass[twocolumn,showpacs,preprintnumbers,amsmath,amssymb,superscriptaddress,prb]{revtex4-1}
\usepackage{graphicx}
\usepackage{dcolumn}
\usepackage{bm}

\begin{document}

\title{
Edge states of zigzag graphene nanoribbons 
with B and N doping at edge atoms
}

\author{Tomoaki Kaneko}
\email{KANEKO.Tomoaki@nims.go.jp}
\affiliation{Computational Materials Science Unit, NIMS, Sengen 1-2-1, Tsukuba 305-0047, Japan}

\author{Kikuo Harigaya}
\email{k.harigaya@aist.go.jp}
\affiliation{Nanosystem Research Institute, AIST, Higashi 1-1-1, Tsukuba 305-8565, Japan}

\date{\today}

\begin{abstract}
Using a tight binding model, we theoretically study the electronic properties of zigzag boron-carbon-nitride (BCN) nanoribbons where the outermost C atoms of zigzag graphene nanoribbons are replaced with B and N atoms.
We show that the flat bands and edge states appear at the Fermi level when the number of B and N atoms are equal except the case of six times supercell.
To investigate the origin of the edge states in BCN nanoribbons, we consider the semi-infinite graphene sheet with zigzag edge where outermost C atoms are replaced with B and N atoms.
We analytically demonstrate the presence of edge states at the zigzag edges and flat bands using a transfer matrix method.
\end{abstract}


\maketitle

\section{Introduction}

Graphene nanoribbons (GNRs) are the stripes of graphene, which were fabricated after cutting graphene sheet by e-beam lithography\cite{Han2007prl,Chen2007physicaE}, solution dispersion and sonication\cite{Li2008Science} , unzipping carbon nanotubes\cite{Jiao2009nature,Kosynkin2009nature}, and bottom-up process\cite{Cai2010nature}.
The electronic and magnetic properties of GNRs were investigated by many authors.\cite{Fujita1996jpsj,Nakada1996prb,Wakabayashi1999prb,Miyamoto1999prb,Son2006prl,Brey2006prb}
Especially, the presence of partially flat bands and peculiar edge states in GNR with zigzag edges were predicted.\cite{Fujita1996jpsj,Nakada1996prb}
Quite recently, observation of the edge states at the zigzag edges in GNR were reported by Tao {\it et al.}\cite{Tao2011nphys}
The electronic transport properties of GNRs also were measured by many groups and the formation of transport gaps were reposted .\cite{Han2007prl,Chen2007physicaE,Li2008Science,Ozyilmaz2007apl,Stampefer2009prl,Gallagher2010prb,Bai2010nnano,Han2010prl,Nakaharai2012apex} 
Therefore, GNRs attract much interest for both scientific and application point of view, because of their outstanding electronic properties.

On the other hand, hexagonal boron nitride (BN) has a similar lattice structure, but it has wide band gap.\cite{Catellani1987prb,Blase1995prb}
Since B and N atoms act as donors and acceptors in graphene, the nanoribbons with B and N atoms, i.e., boron-carbon-nitride (BCN) nanoribbons have potential application to electronic devices with tunable electronic and magnetic properties.
Since BCN nanotubes were synthesized by many groups,\cite{WengSieh1995prb,Redlich1996cpl,Sen1998cpl,Yu2000cpl,Yu2000prl} BCN nanoribbons can be fabricated by unzipping of the BCN nanotubes.
Synthesis of hybridized BN and graphene sheet using thermal catalytic chemical vapor deposition was reported Ci {\it et al}.\cite{Ci2010nmat}
Quite recently, fabrication of BN and graphene in-plane heterostructure with controllable domain shapes were reported.\cite{Liu2013nnano}
BCN nanoribbons can also be fabricated by cutting BCN sheets using e-beam lithography.

The electronic and magnetic properties of BCN nanoribbons were investigated by several authors.\cite{Nakamura2005prb,Kan2008jcp,He2010apl,Basheer2011njp,Lu2011jpcc,Lu2010apl,Xu2010prb,Lai2011Nanoscale,Kaneko2012submitted-GiantStark,Kaneko2012submitted-Alternation,Kaneko2013jpsj}
The ferrimagnetic or half-metallic properties of BCN nanoribbons have been reported. \cite{Nakamura2005prb,Kan2008jcp,He2010apl,Basheer2011njp,Lu2011jpcc,Lu2010apl,Kaneko2013jpsj} 
Appearance of the so-called nearly free electron states just above the Fermi level due to the giant Stark effect by internal electric field was predicted in zigzag BCN nanoribbons where the outermost C atoms of GNR are uniformly replaced with B and N atoms.\cite{Kaneko2012submitted-GiantStark}
For zigzag BCN nanoribbons where B and N atoms are doped around the edges, the flat bands and corresponding edge states were believed to vanish due to different strong modulation of site energies by substitutions.

Previously, Kaneko {\it et al}.\ investigated the electronic and magnetic properties of zigzag BCN nanoribbons where the outermost C atoms are replaced with B and N atoms alternately and showed that the edge states and the flat bands exist at the Fermi level in rich and poor H$_2$ environment.\cite{Kaneko2012submitted-Alternation}
Such flat bands can be found in BC$_2$N nanoribbons with zigzag edges where the atoms are arranged B-C-N-C along the zigzag lines.\cite{Kaneko2013jpsj}
The distributions of charge and spin densities of the edge states in such BCN nanaoribbons are different from those of conventional edge states in GNRs.\cite{Kaneko2012submitted-Alternation,Kaneko2013jpsj}
In the previous paper, however, the origin of the flat bands and edge states were not discussed in detail.
The purpose of present paper is to clarify the origin of edge states and flat bands in such BCN nanoribbons.

In this paper, we theoretically study the flat bands and the edge state in BCN nanoribbons with zigzag edges where outermost C atoms are replaced with B and N atoms by means of a tight binding model.
The band structures of BCN nanoribbons are calculated numerically.
We show that the edge states at the semi infinite BCN sheet are present when the number of B atoms equals to that of N atoms.
We analytically demonstrate the presence of edge states and flat bands.

This paper is organized as follows:
In Sec.\ II, the tight binding model is described.
The numerical results of the band structures of BCN nanoribbons are presented in Sec.\ III.
In Sec.\ IV, the edge states are discussed analytically using a transfer matrix method, and the outlines of demonstrations are presented.
The cell size dependence on the band structures and disappearance of the flat bands and edge states are discussed in Sec.\ V.
A short summary is given in Sec.\ VI.
In Appendix A, the details of demonstrations are summarized.

\section{Tight binding model}

The Hamiltonian of the system within the tight binding model of $\pi$-electrons is given by
\begin{equation}
{\cal H} = \sum_{i} E_i c_{i}^\dagger c_{i} - \sum_{\langle i,j \rangle} t_{i,j} c_{i}^\dagger c_{j},
\end{equation}
where $E_i$ is an energy of $\pi$ electron at the site $i$, $c_{i}^\dag$ and $c_{i}$ are the creation and annihilation operators of electrons at the lattice site $i$, respectively, $\langle i,j\rangle$ stands for summation over the adjacent atoms and $t_{i,j}$ is the hopping integral of $\pi$ electrons from $j$th atom to $i$th atom.
$E_i$ are $E_{\rm B}$, $E_{\rm C}$ and $E_{\rm N}$, the site energies at the B, C and N sites, respectively.
Following to the Yoshioka {\it et al}., we shall assume that the hopping integrals are constant regardless of the atoms, i.e., $t_{i,j}\equiv t$, and $E_{\rm N}=-E_{\rm B}$. \cite{Yoshioka2003jpsj}
For the numerical calculations, we shall choose $E_{\rm B}=t$, $E_{\rm C}=0$, and $E_{\rm N}=-t$. \cite{Yoshioka2003jpsj}
We confirmed that the results does not essentially depend on the choice of the site energies.

\begin{figure}[t!]
\centering
\includegraphics[width=7cm]{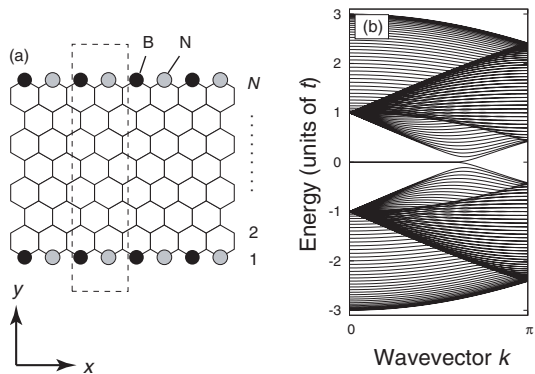}
\caption{
(a) Schematic illustration of B$_\vee$N structure BCN nanoribbon.
In this figure, B and N atoms are indicated by filled and shaded circles, respectively. 
The dashed rectangle represents the unit cell.
(b) The band structure of B$_\vee$N structure BCN nanoribbon for $N=40$.
}
\label{fg:L2-band}
\end{figure}

\section{The band structures of BCN nanoribbons}

In this section, we shall consider the band structure of zigzag BCN nanoribbons where outermost C atoms of GNR are replaced with B and N atoms.
Let $L$ be a size of unit cell in the one-dimensional direction and $N$ be a number of zigzag lines in the nanoribbon.
First, we consider $L=2$ case and the structure of BCN nanoribbons are schematically shown in Fig.\ \ref{fg:L2-band} (a).\cite{Kaneko2012submitted-Alternation}
We shall call this nanoribbon as the B$_\vee$N structure.
In this figure, B and N atoms are indicated by the filled and shaded circles, respectively, and C atoms are located at the empty vertices of the hexagons.
In Fig.\ \ref{fg:L2-band} (b) we show the band structure of such the B$_\vee$N structure nanoribbon with $N=40$.
We clearly see the partially flat bands at $E=0$ in $|k|<\pi/3$ region in the wavevector space.\cite{Kaneko2012submitted-Alternation}
Since the partially flat bands of zigzag GNR are present in $2\pi/3 < |k|$, the band structure of B$_\vee$N structure nanoribbons seemed to be folded from that of GNR.

\begin{figure}[t!]
\centering
\includegraphics[width=7cm]{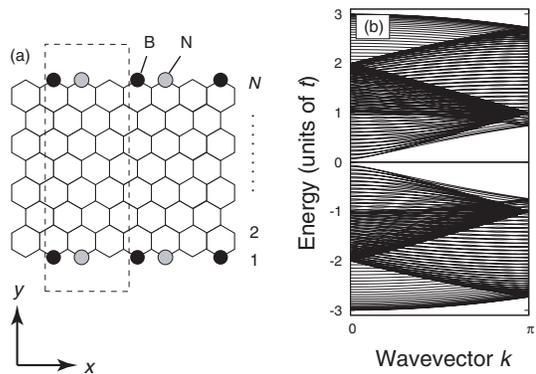}
\caption{
(a) Schematic illustration of B$_\vee$N$_\vee$C structure BCN nanoribbon.
(b) The band structure of B$_\vee$N$_\vee$C structure BCN nanoribbon for $N=40$.
}
\label{fg:L3-band}
\end{figure}

Next, we shall consider $L=3$ case whose structure is schematically shown in Fig.\ \ref{fg:L3-band} (a). 
We shall call this nanoribbon as the B$_\vee$N$_\vee$C structure.
The band structure of B$_\vee$N$_\vee$C structure nanoribbon for $N=40$ is presented in Fig.\ \ref{fg:L3-band} (b). 
We observed the flat bands at $E=0$ in the whole Brillouin zone.
It should be emphasized that we could not obtained the flat bands in the different configurations, such as the B$_\vee$B$_\vee$N or B$_\vee$N$_\vee$N structures.
Since the Brillouin zone is folded into thirds for $L=3$, the band structure is also seemed to be folded from that of GNR.

\begin{figure}[t!]
\centering
\includegraphics[width=8cm]{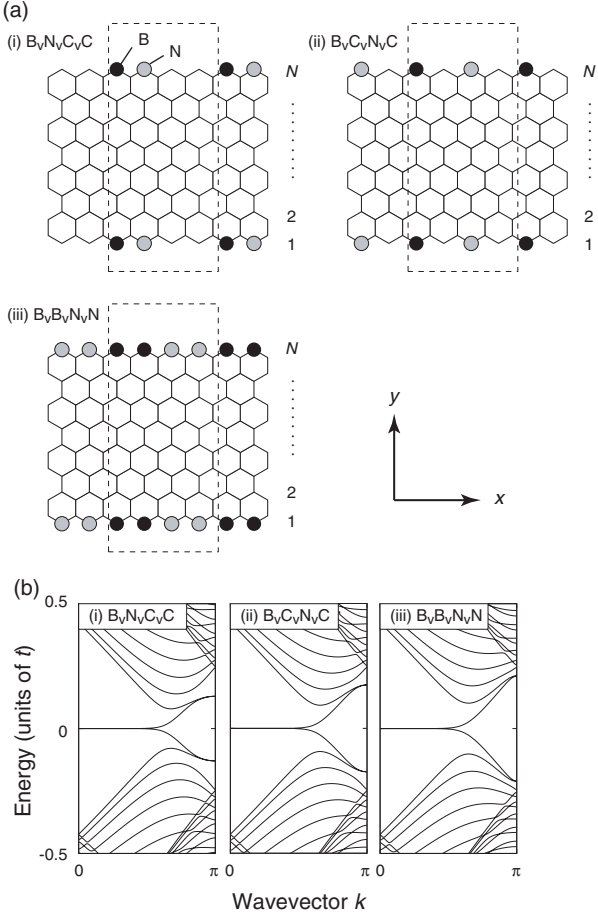}
\caption{
(a) Schematic illustration of BCN nanoribbons for $L=4$ and for B$_\vee$N$_\vee$C$_\vee$C structure (i), B$_\vee$C$_\vee$N$_\vee$C structure (ii) and B$_\vee$B$_\vee$N$_\vee$N structure (iii).
(b) The band structures of BCN nanoribbons for $L=4$ and for B$_\vee$N$_\vee$C$_\vee$C structure (i), B$_\vee$C$_\vee$N$_\vee$C structure (ii) and B$_\vee$B$_\vee$N$_\vee$N structure (iii). 
}
\label{fg:L4-band}
\end{figure}

Next, we move onto $L=4$ case.
Here, we shall consider three different configuration shown in Fig.\ \ref{fg:L4-band} (a).
In the following, we shall call these nanoribbons as B$_\vee$N$_\vee$C$_\vee$C (i),  B$_\vee$C$_\vee$N$_\vee$C (ii) and B$_\vee$B$_\vee$N$_\vee$N (iii) structures.
The band structures of these nanoribbons are presented in Fig.\ \ref{fg:L4-band} (b).
We observed the partially flat bands $|k|<2\pi/3$.
Since the doubly degenerate flat bands exist $|k|<2\pi/3$ and the quadruply degenerate flat bands exist $|k|>2\pi/3$ for GNR with $L=4$,  the flat bands shown in Fig.\ \ref{fg:L4-band} (b) are shorter than those in GNR.
As discussed later, the length of flat bands seems to be related with the position of the K point of the honeycomb lattice and the X point of $L=1$ in the one dimensional Brillouin zone, but such interpretation is not applicable.

Above numerical results suggest that the number of substituted B and N atoms should be equal to obtain the flat bands.
In the next section, we shall consider the edge states at semi infinite BCN sheet in order to investigate the origin of the flat bands and edge states.

\section{Edge States in Semi-infinite Sheet}

\subsection{Transfer matrix}

\begin{figure}[t!]
\centering
\includegraphics[width=7cm]{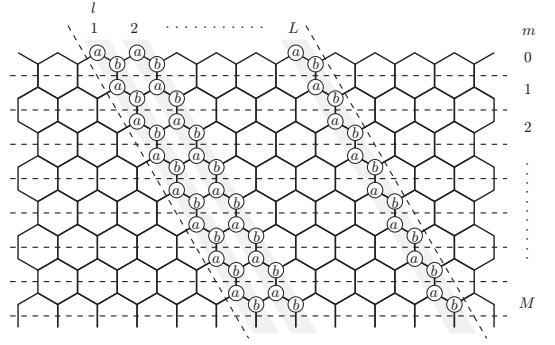}
\caption{
Schematic illustration of a structure of zigzag edge of semi-infinite graphene.
The unit cell is indicated by the dashed lines.
}
\label{fg:semi-inf}
\end{figure}

In Ref.\ [\cite{Wakabayashi2010jpsj}], Wakabayashi {\it et al}.\ investigated the edge states at an alternately beard zigzag edge and a coved edge using a transfer matrix method.
Following the Wakabayashi {\it et al}, we shall construct the transfer matrix method within the tight binding models to investigate the origin of  edge states for arbitrary $L$.
We shall consider semi infinite sheet with zigzag edge schematically shown in Fig.\ \ref{fg:semi-inf}.
In the following, energies are measured in the unit of $t$ and length is measured by $a_0$, and the sublattices are denoted by uncapitalized characters, $a$ and $b$.

Let $\psi_{(a,m,l)}$ and $\psi_{(b,m,l)}$ be the amplitudes at the $(a,m,l)$ and $(b,m,l)$ sites defined in Fig.\ \ref{fg:semi-inf}.
For $E=0$ states, Schr\"odinger equation are given by
\begin{eqnarray}
-\eta^{-1}\psi_{(a,m,1)}-\eta\psi_{(a,m,2)}-\psi_{(a,m+1,1)}&=&0 \nonumber \\
-\eta^{-1}\psi_{(a,m,2)}-\eta\psi_{(a,m,3)}-\psi_{(a,m+1,2)}&=&0 \nonumber \\
\vdots\qquad\qquad\qquad\qquad && \nonumber \\
-\eta^{-1}\psi_{(a,m,L)}-\eta\psi_{(a,m,1)}-\psi_{(a,m+1,L)}&=&0,
\label{eq:asub-eq}
\end{eqnarray}
and
\begin{eqnarray}
-\eta^{-1}\psi_{(b,m,1)}-\eta\psi_{(b,m,2)}-\psi_{(b,m-1,2)}&=&0 \nonumber \\
-\eta^{-1}\psi_{(b,m,2)}-\eta\psi_{(b,m,3)}-\psi_{(b,m-1,3)}&=&0 \nonumber \\
\vdots\qquad\qquad\qquad\qquad && \nonumber \\
-\eta^{-1}\psi_{(b,m,L)}-\eta\psi_{(b,m,1)}-\psi_{(b,m-1,1)}&=&0,
\label{eq:bsub-eq}
\end{eqnarray}
where $\eta=\exp(ik/2L)$.
For the outermost atoms, the amplitudes satisfy the initial conditions
\begin{eqnarray}
-\eta^{-1}\psi_{(b,0,1)}-\eta\psi_{(b,0,2)}+ E_2\psi_{(a,0,2)}&=&0 \nonumber \\
-\eta^{-1}\psi_{(b,0,2)}-\eta\psi_{(b,0,3)}+ E_3\psi_{(a,0,3)}&=&0 \nonumber \\
\vdots\qquad\qquad\qquad\qquad && \nonumber \\
-\eta^{-1}\psi_{(b,0,L)}-\eta\psi_{(b,0,1)}+ E_1\psi_{(a,0,1)}&=&0,
\label{eq:initial-cond}
\end{eqnarray}
where $E_l$ ($l=1,\cdots,L$) are the site energies at $(a,0,l)$ sites and take $E_{\rm B}$, $E_{\rm C}$ and $E_{\rm N}$.
The amplitudes at the $a$-sublattice sites, $\psi_{(a,0,l)}$ $(l=1,\cdots,L)$, are same as those in conventional zigzag edges.
On the other hand, $\psi_{(b,0,l)}$ $(l=1,\cdots,L)$ can be obtained from $\psi_{(a,0,l)}$, $(l=1,\cdots,L)$ by solving Eq.\ (\ref{eq:initial-cond}).
Therefore, the amplitudes at $b$-sublattice sites become finite, while those of  the edge states at the conventional zigzag edges vanish identically.
The amplitudes at $b$-sublattice sites play decisive role for the formation of the edge states at the BCN edges as discussed below.

\begin{figure*}[t!]
\centering
\includegraphics[width=12cm]{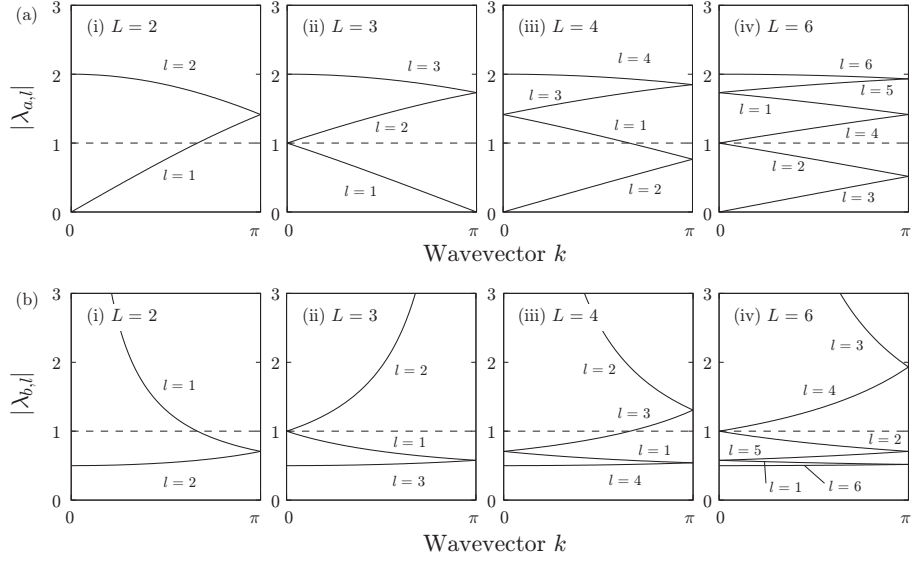}
\caption{
(a) The absolute values of eigenvalues of $T_a$, $\lambda_{a,l}$, as functions of wavevector, $k$, for $L=2$ (i), $L=3$ (ii), $L=4$ (iii) and $L=6$ (iv).  
(b) Same figures for $\lambda_{b,l}$.
}
\label{fg:lambda-2346}
\end{figure*}

Let $\bm{\psi}_{a,m}$ and $\bm{\psi}_{b,m}$ be the column vectors of amplitudes at $a$- and $b$-sublattice in $m$th low defined by
\begin{equation}
\bm{\psi}_{a,m}=\left(
\begin{array}{c}
\psi_{(a,m,1)} \\
\psi_{(a,m,2)} \\
\vdots \\
\psi_{(a,m,L)} \\
\end{array}
\right),
\end{equation}
and 
\begin{equation}
\bm{\psi}_{b,m}=\left(
\begin{array}{c}
\psi_{(b,m,1)} \\
\psi_{(b,m,2)} \\
\vdots \\
\psi_{(b,m,L)} \\
\end{array}
\right),
\end{equation}
respectively.
Then, Eqs.\ (\ref{eq:asub-eq}) and (\ref{eq:bsub-eq}) can be summarized as
\begin{equation}
\bm{\psi}_{a,m} = T_a^m \bm{\psi}_{a,0},
\end{equation}
and 
\begin{equation}
\bm{\psi}_{b,m} = T_b^m \bm{\psi}_{b,0},
\end{equation}
where 
\begin{equation}
T_a=\left(
\begin{array}{ccccc}
-\eta^{-1}&-\eta &&& \\
&-\eta^{-1}&-\eta && \\
&&-\eta^{-1}&\ddots & \\
&&&\ddots&-\eta  \\
-\eta &&&&-\eta^{-1} \\
\end{array}
\right),
\end{equation}
and
\begin{equation}
T_b =\left(
\begin{array}{ccccc}
-\eta &&&&-\eta^{-1} \\
-\eta^{-1}&-\eta &&& \\
&-\eta^{-1}&-\eta && \\
&&\ddots   &\ddots & \\
&&&-\eta^{-1}&-\eta
\end{array}
\right)^{-1}.
\end{equation}
are the transfer matrices.
Note that $T_b$ is defined by the inverse matrix.
Using $T_b$, the initial condition Eq.\ (\ref{eq:initial-cond}) can be written as 
\begin{equation}
\bm{\psi}_{b,0}=T_bE_{\rm diag}\bm{\psi}_{a,0}, 
\label{eq:ini-matrix}
\end{equation}
where $E_{\rm diag}$ is a diagonal matrix with the elements $E_1, E_2, \cdots, E_L$ in this order.

The $m$th power of the transfer matrices can be written as
\begin{equation}
T_a^m = \sum_{l=1}^L \lambda_{a,l}^m \bm{C}_{a,l}\bm{C}_{a,l}^\dag,
\end{equation}
and
\begin{equation}
T_b^m = \sum_{l=1}^L \lambda_{b,l}^m \bm{C}_{b,l}\bm{C}_{b,l}^\dag,
\end{equation}
where the eigenvalues of $T_a$ and $T_b$ are 
\begin{equation}
\lambda_{a,l} = -\eta^{-1} - \eta e^{2\pi i l/L},
\end{equation}
and
\begin{equation}
\lambda_{b,l} = \frac{-1}{\eta + \eta^{-1} e^{2\pi i l/L}},
\end{equation}
respectively, and the eigenvectors are 
\begin{equation}
\bm{C}_{a,l} = \frac{1}{\sqrt{L}}
\left(
\begin{array}{c}
1\\
e^{2\pi i l/L}\\
\vdots \\
e^{2\pi i l(L-1)/L}
\end{array}
\right),
\end{equation}
and
\begin{equation}
\bm{C}_{b,l} = \frac{1}{\sqrt{L}}
\left(
\begin{array}{c}
e^{2\pi i l(L-1)/L}\\
\vdots \\
e^{2\pi i l/L}\\
1
\end{array}
\right),
\end{equation}
respectively.

In Fig.\ \ref{fg:lambda-2346}, the absolute values of eigenvalues are plotted as functions of wavevector, $k$.
When the absolute values of eigenvalues of the transfer matrices are larger than unity, these states does not satisfy the boundary condition.
The edge states are defined as evanescent states at the edge, i.e., the edge states exist if the absolute values of eigenvalues, $\lambda_{a,l}$ and $\lambda_{b,l'}$ ($l,l'=1,\cdots,L$), are smaller than unity. 
Therefore, investigating the behavior of these amplitudes with increasing of $m$, we can demonstrate the presence of edge states.

The procedure of the demonstration can be summarized as follow:
From Fig.\ \ref{fg:lambda-2346} (a), we can determine $\bm{\psi}_{a,0}$.
Then, we obtain
\begin{equation}
\bm{\psi}_{a,m}=\lambda_{a,l}^m \bm{C}_{a,l}.
\label{eq:aM-amplitude}
\end{equation}
Using Eq.\ \ref{eq:ini-matrix}, we can write
\begin{equation}
\bm{\psi}_{b,m}=\sum_{l'=1}^L \lambda_{b,l'}^m (\bm{C}_{b,l'},T_bE_{\rm diag}\bm{C}_{a,l})\bm{C}_{b,l'}.
\label{eq:bM-amplitude}
\end{equation}
Since this summation includes the eigenvalues, $\lambda_{b,l'}$, whose absolute values are larger than unity, the state does not satisfy the boundary condition.
Therefore, the inner products, $(\bm{C}_{b,l'},T_b E_{\rm diag}\bm{C}_{a,l})$, must vanish when the absolute values of eigenvalues are larger than unity in order to have the edge states.
The evaluation of these inner products plays decisive role for the demonstration of the edge states.

Here, we shall consider the condition of the appearance of edge states for arbitrary $L$.
The absolute values of eigenvalues can be written as $|\lambda_{a,l}|=2|\cos(k/2L + \pi l/L)|$ and $|\lambda_{b,l}|=1/2|\cos(k/2L - \pi l/L)|$.
For even $L$, $\lambda_{a,L/2}$ vanishes and $\lambda_{b,L/2}$ diverges at $k=0$.
For odd $L$,  $\lambda_{a,(L-1)/2}$ vanishes and $\lambda_{b,(L+1)/2}$ diverges at $k=\pi$.
Therefore, we need to evaluate the inner products $(\bm{C}_{b,L/2},T_b E_{\rm diag}\bm{C}_{a,L/2})$ for even $L$ and $(\bm{C}_{b,(L+1)/2},T_b E_{\rm diag}\bm{C}_{a,(L-1)/2})$ for odd $L$.
As demonstrated in Appendix A-1, these inner products are proportional to $\sum_{l=1}^L E_l$ which vanishes when the number of B and N atoms are equal due to the definition of site energies.
Therefore, we have shown that the number of B and N atoms must be equal in order to have the edge states.
It should be emphasized that this condition is independent of the atomic arrangement.
In the following, we apply these results for $L=2,3$ and 4 cases and investigate the properties of edge states and flat bands.

\subsection{$L=2$ case}

First, we shall consider $L=2$ case.
Since $|\lambda_{a,1}|<1$ for $0<k<2\pi/3$ as shown in Fig.\ \ref{fg:lambda-2346} (a)-(i), the amplitude at the outermost sites are given by
\begin{equation}
\bm{\psi}_{a,0} = \bm{C}_{a,1} = \frac{1}{\sqrt{2}}
\left(
\begin{array}{c}
1 \\ -1
\end{array}
\right).
\end{equation}
Then, the amplitudes at $a$-sublattice sites in the $m$th low can be written as $\bm{\psi}_{a,m} = \lambda_{a,1}^m \bm{C}_{a,1}$.\cite{Kaneko2012submitted-Alternation}
Using Eq.\ (\ref{eq:initial-cond2}), we obtain
\begin{eqnarray}
\bm{\psi}_{b,0} &=& T_b E_{\rm diag} \bm{C}_{a,1} \nonumber \\
&=&\frac{1}{\sqrt{2}(\eta^{-2}-\eta^2)}
\left(
\begin{array}{c}
-\eta^{-1} E_2 -\eta E_1 \\
 \eta^{-1} E_1 +\eta E_2
\end{array}
\right). 
\end{eqnarray}
Since $|\lambda_{b,1}|>1$ for $0<k<2\pi/3$, the inner product,
\begin{equation}
(\bm{C}_{b,2},T_b E_{\rm diag} \bm{C}_{a,1}) = \frac{E_1+E_2}{2(\eta^{-1}-\eta)},
\end{equation}
must vanish identically in order to form the edge states.
Then, we obtain $E_1+E_2=0$, i.e., the B$_\vee$N structure is allowed to have the edge states.
For the the B$_\vee$N structure, we obtain
\begin{equation}
\bm{\psi}_{b,0} = \frac{E_{\rm B}}{\sqrt{2}(\eta + \eta^{-1})}
\left(
\begin{array}{c}
1 \\ 1
\end{array}
\right)
= \frac{E_{\rm B}}{\eta + \eta^{-1}}\bm{C}_{b,2}.
\end{equation}
Therefore, the amplitude at the $b$-sublattice in the $m$th low can be written as\cite{Kaneko2012submitted-Alternation}
\begin{equation}
\bm{\psi}_{b,m} = \frac{E_{\rm B} \lambda_{b,2}^m}{\eta + \eta^{-1}} \bm{C}_{b,2}.
\end{equation}
As shown in Fig.\ \ref{fg:lambda-2346} (a) $|\lambda_{b,1}| \leq 1$ in $0 \leq k \leq \pi$, the flat bands exist in $0 \leq k < 2\pi/3$, showing agreement with numerical results as shown above and the results within the first-principles calculations.\cite{Kaneko2012submitted-Alternation}

\subsection{$L=3$ case}

Next, we shall consider the $L=3$ case.
Since the procedure of the proof is similar to the case of $L=2$, we present the outline of the proof.
The details of calculations are summarized in Appendix A-2.

Since $|\lambda_{a,l}|<1$ for $l=1$, the amplitude at the $a$-sublattice is given by $\bm{\psi}_{a,m}=\lambda_{a,1}^m  \bm{C}_{a,1}$.
Then, the amplitude at the $b$-sublattice in the $m$th low is given by Eq.\ (\ref{eq:bM-amplitude}).
Since $|\lambda_{b,2}|>1$ as shown in Fig.\ \ref{fg:lambda-2346} (b)-(ii), the inner product $(\bm{C}_{b,2},T_b E_{\rm diag}\bm{C}_{a,1})$ must vanish in order to have the edge states.
As presented in Appendix A-2,
\begin{equation}
(\bm{C}_{b,2},T_b E_{\rm diag}\bm{C}_{a,1})\propto E_1+E_2+E_3.
\end{equation}
Therefore, the number of B atoms and N atoms must be equal, i.e., the B$_\vee$N$_\vee$C structure is allowed to have the edge states.

\begin{figure*}[t!]
\centering
\includegraphics[width=15cm]{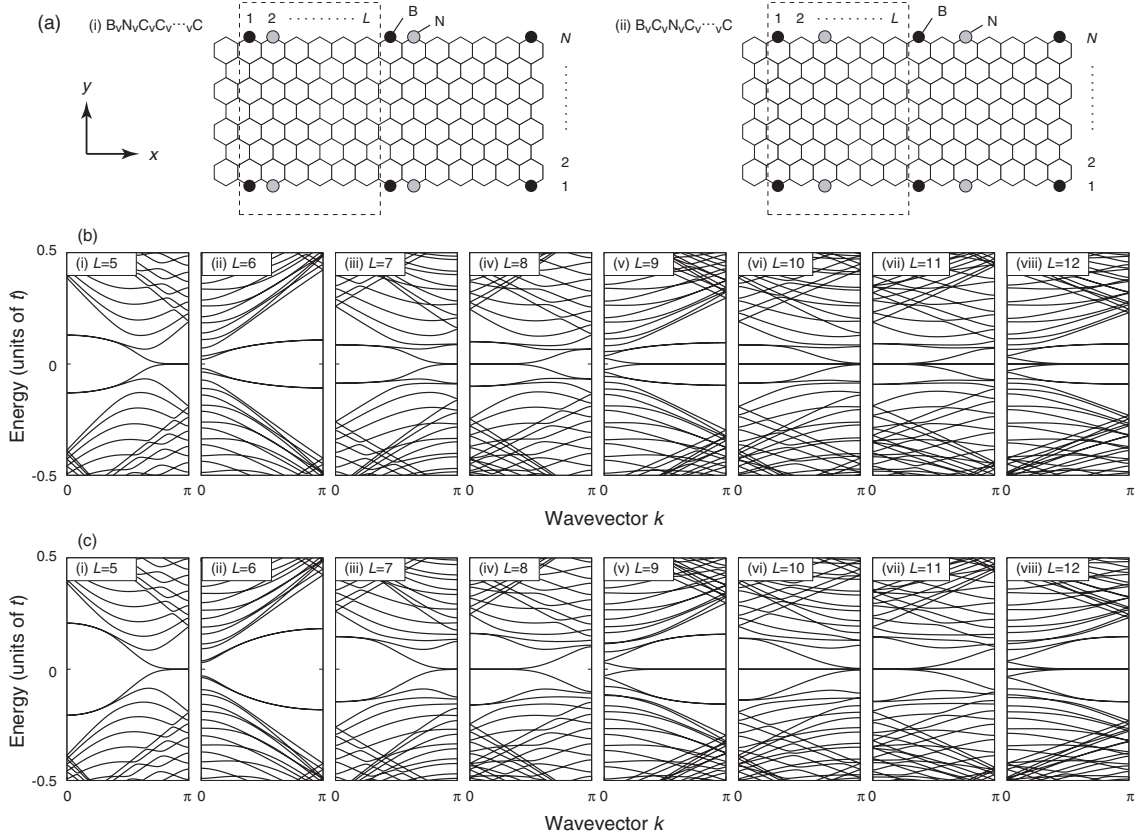}
\caption{
(a) Schematic illustration of BCN nanoribbons with B$_\vee$N$_\vee$C$_\vee$C$_\vee\cdots_\vee$C (i) and B$_\vee$C$_\vee$N$_\vee$C$_\vee\cdots_\vee$C (ii) structures.
(b) The band structure of BCN nanoribbons for $N=40$ with B$_\vee$N$_\vee$C$_\vee\cdots_\vee$C structures.
(i) for $L=5$, (ii) for $L=6$,  (iii) for $L=7$, (iv) for $L=8$, (v) for $L=9$, (vi) for $L=10$, (vii) for $L=11$ and (viii) for $L=12$.
For $L\geq9$, the flat bands cover the whole Brillouin zone even when the K$_0$ and X$_1$ points locate at the same point.
(c) Same figure as (b) for B$_\vee$C$_\vee$N$_\vee$C$_\vee\cdots_\vee$C (ii) structure.
The length of the flat bands is independent of the atomic arrangement.
}
\label{fg:L-dep}
\end{figure*}

\subsection{$L=4$ case}

Next, we shall consider the $L=4$ case.
Here, we also present the online of the demonstration and the detail are summarized in Appendix A-3.
From Fig.\ \ref{fg:lambda-2346} (a)-(iii), we can write $\bm{\psi}_{a,m}=\lambda_{a,1}^m\bm{C}_{a,1}$ for $2\pi/3<k<\pi$ and $\bm{\psi}_{a,m}=\lambda_{a,2}^m\bm{C}_{a,2}$ for the whole Brillouin zone.
On the other hand, $|\lambda_{b,2}|>1$ for the whole Brillouin zone and $|\lambda_{b,3}|>1$ for $2\pi/3<k<\pi$.
Therefore, we have to evaluate the inner product $(\bm{C}_{b,l'},T_b E_{\rm diag}\bm{C}_{a,l})$ for $l=1$ and 2, and for $l'=2$ and 3.
As presented in Appendix A-2, we obtain
\begin{eqnarray}
(\bm{C}_{b,2},T_b E_{\rm diag}\bm{C}_{a,1}) &\propto& E_1-iE_2-E_3+iE_4, \label{eq:L4-1}\\
(\bm{C}_{b,3},T_b E_{\rm diag}\bm{C}_{a,1}) &\propto& E_1+E_2+E_3+E_4, \label{eq:L4-2}\\
(\bm{C}_{b,2},T_b E_{\rm diag}\bm{C}_{a,2}) &\propto& E_1+E_2+E_3+E_4, \label{eq:L4-3}\\
\mbox{and} \qquad \qquad \qquad \qquad && \nonumber \\
(\bm{C}_{b,3},T_b E_{\rm diag}\bm{C}_{a,2}) &\propto& E_1+iE_2-E_3-iE_4. \label{eq:L4-4}
\end{eqnarray}
Therefore, Eqs.\ (\ref{eq:L4-2}) and (\ref{eq:L4-3}) give
\begin{equation}
 E_1+E_2+E_3+E_4=0,
 \label{eq:cond-L4-1}
\end{equation}
and Eqs.\ (\ref{eq:L4-1}) and (\ref{eq:L4-4}) give
\begin{equation}
E_1-E_3 = E_2-E_4=0.
 \label{eq:cond-L4-2}
 \end{equation}
It should be noted that the B$_\vee$N$_\vee$B$_\vee$N structure satisfies both Eqs.\ (\ref{eq:cond-L4-1})  and (\ref{eq:cond-L4-2}).
On the other hand, the B$_\vee$N$_\vee$C$_\vee$C, B$_\vee$C$_\vee$N$_\vee$C and B$_\vee$B$_\vee$N$_\vee$N structures satisfy Eq.\ (\ref{eq:cond-L4-1}).

Since  the B$_\vee$N$_\vee$C$_\vee$C, B$_\vee$C$_\vee$N$_\vee$C and B$_\vee$B$_\vee$N$_\vee$N structures do not satify Eqs.\  (\ref{eq:L4-1}) and (\ref{eq:L4-2}) simultaneously, $\bm{\psi}_{a,0}$ did not have the $\bm{C}_{a,1}$ components.
Therefore, the amplitudes can be written as $\bm{\psi}_{a,0} = \bm{C}_{a,2}$ for $0<k<2\pi/3$.
This gives the flat bands range of $k$ for $L=4$ and shows good agreement with the numerical results shown in Fig.\ \ref{fg:L4-band} (b).
For  the B$_\vee$N$_\vee$B$_\vee$N structure, $\bm{\psi}_{a,0} = \bm{C}_{a,1}$ for $0<k<2\pi/3$ and $\bm{\psi}_{a,0} = \bm{C}_{a,2}$ for the whole Brillouin zone.
This result agrees with the numerical results shown in Fig.\ \ref{fg:L2-band} (b).

\section{Discussion}

\begin{table}[b!]
\centering
\caption{
The cell size dependence of the special points of the Brillouin zone. 
K$_0$ and X$_1$ stand for the K point of honeycomb lattice and the X point for $L=1$ nanoribbon, respectively.
$\Delta k$ is the distance between K$_0$ and X$_1$.
}
\begin{tabular}{cccc|cccccccc}
\hline \hline 
&\quad&$L$&\quad&\quad& K$_0$ &\quad& X$_1$ &\quad& $\Delta k$ &\quad&\\
\hline
&&2 && &$2\pi/3$ && 0     && $2\pi/3$ &\\
&&3 && &0        && $\pi$ && $\pi$    &\\
&&4 && &$2\pi/3$ && 0     && $2\pi/3$ &\\
&&5 && &$2\pi/3$ && $\pi$ && $\pi/3$  &\\
&&6 && &0        && 0     && 0        &\\
&&7 && &$2\pi/3$ && $\pi$ && $\pi/3$  &\\
&&8 && &$2\pi/3$ && 0     && $2\pi/3$ &\\
&&9 && &0        && $\pi$ && $\pi$  &\\
\hline \hline 
\end{tabular}
\end{table}

In the previous section, we have analytically investigated the edge states and flat bands at zigzag BCN edge for $L=2$, 3 and 4 cases.
We found that the condition ``the number of B and N atoms are equal" ensures the disappearance of divergent component at the $b$-sublattice.
However, there are exceptions which we shall discuss in this section.

As shown in Figs.\ \ref{fg:L-dep} (a)-(i) and (ii), we shall consider BCN nanoribbons with the B$_\vee$N$_\vee$C$_\vee$C$_\vee\cdots_\vee$C and  B$_\vee$C$_\vee$N$_\vee$C$_\vee\cdots_\vee$C structures since we confirmed that the length of the flat bands does not depend on the arrangement of the B and N atoms.
Figure \ref{fg:L-dep} (b) shows the band structure of BCN nanoribbons for $N=40$ with B$_\vee$N$_\vee$C$_\vee\cdots_\vee$C structures.
As shown in Fig.\ \ref{fg:L-dep} (b), we observed the flat bands with the length $\pi/3$ for $L=5$ and 7, $2\pi/3$ for $L=8$ and $\pi$ for $L\geq 9$ nanoribbons.
However, the flat bands vanish for $L=6$ nanoribbons.
The band structure of BCN nanoribbons for $N=40$ with B$_\vee$C$_\vee$N$_\vee$C$_\vee\cdots_\vee$C structures is presented in Fig.\ \ref{fg:L-dep} (c).
As mentioned just above, the length of flat bands is independent of atomic arrangement and the flat bands are absent in BCN nanoribbons for $L=6$.
For $L=6$, we could not observe the flat bands even in the different atomic arrangement except B$_\vee$N$_\vee$C$_\vee$B$_\vee$N$_\vee$C and B$_\vee$N$_\vee$B$_\vee$N$_\vee$B$_\vee$N.
It should be emphasized that the exceptions have smaller unit cells. 
We confirmed that the flat bands appear independent of the atomic arrangement when the number of B and N atoms are equal for arbitrary $L$ except for $L=6$ as demonstrated in the previous section.

Here, we briefly see these effect analytically and the details are summarized in Appendix A-4.
As shown in Appendix A-4, flat bands appear when following two equations are satisfied simultaneously:
\begin{eqnarray}
&& E_1+E_2+E_3+E_4+E_5+E_6= 0,\\
&& E_1-E_4 = E_3-E_6 = E_5- E_2. 
\end{eqnarray}
We find that the B$_\vee$N$_\vee$C$_\vee$B$_\vee$N$_\vee$C and B$_\vee$N$_\vee$B$_\vee$N$_\vee$B$_\vee$N structures satisfy these equation, while the B$_\vee$N$_\vee$C$_\vee$C$_\vee$C$_\vee$C and B$_\vee$C$_\vee$N$_\vee$C$_\vee$C$_\vee$C structures do not satisfy.
Therefore, the flat bands are absent for these two exceptiocal $L=6$ BCN nanoribbons.

Finally, we shall discuss the relation between the length of the flat bands and the special points in the Brillouin zone.
For zigzag GNRs, i.e., $L=1$ case, the flat bands exist from $k=2\pi/3$ to X point  i.e., $k=\pi$.
At  $k=2\pi/3$, a projection of the K and K$'$ point in the Brillouin zone honeycomb lattice to one dimensional Brillouin zone is located.
We shall denote the position of K point in one dimensional Brillouin zone as K$_0$ and X point for $L=1$ as $X_1$.
The position of these points in $L$ times unit cell are summarized in Table 1.
As mentioned above, this behavior seems to be related to the position of the X point in $L=1$ nanoribbon and the K point in the honeycomb lattice.
In this table, $\Delta k$ stands for the distance between the K$_0$ and X$_1$ points.
We found that the calculated results of the length of flat bands accords well with $\Delta k$ for $L=2-9$ structures.
However, the flat bands cover the whole Brillouin zone for $L>9$ structures, but $\Delta k$ oscillates with increasing of $L$.
Therefore, such interpretation is limited only for small $L$ and is not applicable for $L>9$ nanoribbons.

\section{Summary}

In this paper, we theoretically studied the flat bands and the edge states in zigzag BCN nanoribbons where the outermost C atoms of graphene nanoribbons are replaced with B and N atoms using the tight-biding model.
We numerically showed that the flat bands exist at the Fermi level when the number of B and N atoms are equal independent of the atomic arrangement except for the six times supercell.
To investigate the edge states in BCN nanoribbons, we analyzed the edge states at the semi-infinite graphene sheet with zigzag edge where the outermost C atoms are replaced with B and N atoms based on the transfer matrix method for the arbitrary cell size.
Then, we analytically demonstrated that the edge states appear when the numerically obtained condition is satisfied.
For this demonstration, the amplitudes at the sublattice sites which do not belong to the outermost sublattice play decisive role.
The charge distributions of corresponding edge states are different from those in the conventional edge states.
The length of the flat bands seems to be related with the distance of the special point of Brillouin zone, but such interpretation is limited to small cell size regime.

\section*{Acknowledgments}

The authors acknowledge H.\ Imamura, Y.\ Shimoi, H.\ Tsukahara, K.\ Wakabayashi and S.\ Dutta for valuable discussions.
This research was supported by the International Joint Work Program of Daeduck Innopolis under the Ministry of Knowledge Economy (MKE) of the Korean Government.

\appendix

\section{Details of demonstrations}

In this appendix, we shall give the details of demonstrations in the transfer matrix calculations which we omitted above.

\subsection{Arbitrary $L$ case}

Here, we shall present the detail of calculation for arbitrary $L$.
Since, the transfer matrix of $b$-sublattice is given by
\begin{widetext}
\begin{equation}
T_b = \frac{1}{\eta^L-(-1)^L\eta^{-L}}
\left(
\begin{array}{ccccc}
(-1)^{L-1} \eta^{2L-1} & \eta & - \eta ^3 & \cdots  & (-1)^{L-2} \eta^{2L-3}  \\
(-1)^{L-2} \eta^{2L-3}  & (-1)^{L-1} \eta^{2L-1} & \eta & \cdots  & (-1)^{L-3} \eta^{2L-5}  \\
\vdots &\vdots & \ddots & \cdots & \cdots  \\
- \eta^3 & \eta^5 & \cdots &  (-1)^{L-1} \eta^{2L-1} & \eta \\
\eta & - \eta ^3 & \cdots & (-1)^{L-2} \eta^{2L-3} & (-1)^{L-1} \eta^{2L-1} \\
\end{array}
\right),
\end{equation}
\end{widetext}
we obtain 
\begin{widetext}
\begin{eqnarray}
\label{eq:initial-cond2}
\bm{\psi}_{b,0} &=& T_bE_{\rm diag}\bm{\psi}_{a,0} \\
&=&\frac{1}{e^{ik}-(-1)^L}
\left(
\begin{array}{c}
E_2 \psi_{(a,0,2)} \eta - E_3 \psi_{(a,0,3)} \eta ^3 \cdots 
+(-1)^{L-2} E_L \psi_{(a,0,L)} \eta^{2L-3} +(-1)^{L-1} E_1 \psi_{(a,0,1)} \eta^{2L-1} \\
E_3 \psi_{(a,0,3)} \eta - E_4 \psi_{(a,0,4)} \eta ^3 \cdots 
+(-1)^{L-2} E_1 \psi_{(a,0,1)} \eta^{2L-3} +(-1)^{L-1} E_2 \psi_{(a,0,2)} \eta^{2L-1} \\
\vdots\\
E_L \psi_{(a,0,L)} \eta - E_1 \psi_{(a,0,1)} \eta ^3 \cdots 
+(-1)^{L-2} E_{L-2} \psi_{(a,0,L-2)} \eta^{2L-3} +(-1)^{L-1} E_{L-1} \psi_{(a,0,L-1)} \eta^{2L-1} \\
E_1 \psi_{(a,0,1)} \eta - E_2 \psi_{(a,0,2)} \eta ^3 \cdots 
+(-1)^{L-2} E_{L-1} \psi_{(a,0,L-1)} \eta^{2L-3} +(-1)^{L-1} E_L \psi_{(a,0,L)} \eta^{2L-1} \\
\end{array}
\right).\nonumber 
\end{eqnarray}
\end{widetext}

As we discussed above, we need to examine the inner products $(\bm{C}_{b,L/2},T_b E_{\rm diag}\bm{C}_{a,L/2})$ for even $L$ and $(\bm{C}_{b,(L+1)/2},T_b E_{\rm diag}\bm{C}_{a,(L-1)/2})$ for odd $L$.
For even $L$ case, the eigenvectors are given by
\begin{equation}
\bm{C}_{a,L/2} = \frac{1}{\sqrt{L}} \left(
\begin{array}{c}
1 \\ -1 \\ \vdots \\ -1
\end{array}
\right),
\end{equation}
and
\begin{equation}
\bm{C}_{b,L/2} = \frac{1}{\sqrt{L}} \left(
\begin{array}{c}
-1 \\ \vdots \\ -1 \\ 1
\end{array}
\right).
\end{equation}
Then, the inner product can be written as
\begin{equation}
(\bm{C}_{a,L/2},T_bE_{\rm diag}\bm{C}_{a,L/2}) 
= \frac{\eta+\eta^3+\cdots+\eta^{2L-1}}{1-e^{ik}} \sum_{l=1}^L E_l.
\end{equation}
For odd $L$ case, the eigenvectors are given by
\begin{equation}
\bm{C}_{a,(L-1)/2} = \frac{1}{\sqrt{L}} \left(
\begin{array}{c}
1 \\ -\alpha^{-1} \\ \vdots \\ \alpha^{-L+1}
\end{array}
\right),
\end{equation}
and
\begin{equation}
\bm{C}_{b,(L+1)/2} = \frac{1}{\sqrt{L}} \left(
\begin{array}{c}
\alpha^{L-1} \\  \vdots \\ -\alpha\\ 1
\end{array}
\right),
\end{equation}
with $\alpha=e^{\pi i/L}$.
Then,  the inner product can be written as 
\begin{eqnarray}
&&(\bm{C}_{a,(L+1)/2},T_bE_{\rm diag}\bm{C}_{a,(L-1)/2}) \nonumber \\
&=& \frac{\eta+\alpha^{-1}\eta^2+\cdots+\alpha^{-L+1}\eta^{2L-1}}{1+e^{ik}} \sum_{l=1}^L E_l.
\end{eqnarray}
Therefore, these inner products are proportional to the sum of site energies, $\sum_{l=1}^L E_l$.
This quantity vanishes when the number of B and N atoms are equal.

\subsection{$L=3$ case}

Next, we shall consider the case $L=3$.
As shown in Fig.\ \ref{fg:lambda-2346} (a)-(ii), $|\lambda_{a,1}|\leq 1$ but $|\lambda_{a,2}|,|\lambda_{a,3}|\geq 1$ in the whole Brillouin zone.
Therefore, we obtain
\begin{equation}
\bm{\psi}_{a,0} = \bm{C}_{a,1} = \frac{1}{\sqrt 3} \left(
\begin{array}{c}
1 \\ \omega \\ \omega^2
\end{array}
\right),
\end{equation}
with $\omega=e^{2\pi i/3}$ begin a cubic root of unity, and
\begin{equation}
\bm{\psi}_{a,m} = \lambda_{a,1}^m \bm{C}_{a,1}.
\end{equation}
For the $b$-sublattice of 0th low, on the other hand, Eq.\ (\ref{eq:initial-cond}) gives
\begin{equation}
\bm{\psi}_{b,0} = \frac{1}{\sqrt{3}(1+e^{ik})}
\left(
\begin{array}{c}
E_2\omega\eta^1 - E_3\omega^2\eta^3 + E_1\eta^5 \\
E_3\omega^2\eta^1 - E_1\eta^3 + E_2\omega\eta^5 \\
E_1\eta^1 - E_2\omega\eta^3 + E_3\omega^2\eta^5 
\end{array}
\right).
\end{equation}
Then, the amplitudes at the $b$-sublattice sites of $m$th low can be written as
\begin{equation}
\bm{\psi}_{b,m} = \sum_{l=1}^3 \lambda_{b,l}^m \bm{C}_{b,l} (\bm{C}_{b,l},\bm{\psi}_{b,0}).
\end{equation}
As shown in Fig.\ \ref{fg:lambda-2346} (b)-(ii), since $|\lambda_{b,2}|>1$ in the whole Brillouin zone, the inner product, 
\begin{equation}
(\bm{C}_{b,2},T_b E_{\rm diag}\bm{C}_{a,1}) = \frac{(E_1+E_2+E_3)(\omega^2\eta^5-\omega\eta^3+\eta)}{3(1+e^{ik})},
\end{equation}
must vanish to have the edge states  at the BCN edges.
Therefore, the edge states appear when $E_1+E_2+E_3=0$.
In order to satisfy this relation, the number of B atoms and N atoms must be equal.
It should be emphasized that the disappearance of $\bm{C}_{b,2}$ term plays decisive role in this case.
We confirmed that the $\bm{C}_{b,2}$ do not disappear for the other configurations such as B$_\vee$B$_\vee$N or B$_\vee$N$_\vee$N, resulting in the vanishing of the flat bands and edge states as discussed above.

For ${\rm B}_\vee{\rm N}_\vee{\rm C}$ structure, i.e., $E_1=E_{\rm B}$, $E_2=-E_{\rm B}$ and $E_3=0$, there are edge states.
This $\bm{\psi}_{b,0}$ shows $(\bm{C}_{b,2},\bm{\psi}_{b,0})=0$, but $(\bm{C}_{b,l},\bm{\psi}_{b,0})\neq0$ ($l=1$ and 3).
Therefore, the amplitudes at the $b$-sublattice sites can be written as 
\begin{equation}
\bm{\psi}_{b,m} = \lambda_{b,1}^m \bm{C}_{b,1}(\bm{C}_{b,1},\bm{\psi}_{b,0})
+ \lambda_{b,3}^m \bm{C}_{b,3}(\bm{C}_{b,3},\bm{\psi}_{b,0}).
\end{equation}
Due to $|\lambda_{b,1}|\leq 1$ and $|\lambda_{b,3}|\leq 1$, the flat bands exist in the whole Brillouin zone.
This result agrees well with the numerical band structure as shown in Fig.\ 2 (b).

\subsection{$L=4$ case}

Next, we move to the $L=4$ case.
As shown in Fig.\ \ref{fg:lambda-2346} (a)-(iii), $|\lambda_{a,1}|<1$ in $2\pi/3 <k\leq \pi$ and $|\lambda_{a,2}|<1$ in the whole Brillouin zone.
Therefore, there are two possibility for the amplitudes at the outermost $a$-sublattice, i.e.,
\begin{equation}
\bm{\psi}_{a,0} = \bm{C}_{a,1} \quad \mbox{or} \quad \bm{C}_{a,2}.
\end{equation}
On the other hand, since $|\lambda_{b,2}|>1$ in the whole Brillouin zone and $|\lambda_{b,3}|>1$ in $2\pi/3 <k\leq \pi$ shown in Fig.\ \ref{fg:lambda-2346} (b)-(iii), the inner products, $(\bm{C}_{b,2},\bm{\psi}_{b,0})$ and $(\bm{C}_{b,3},\bm{\psi}_{b,0})$, must vanish in order to have the edge states at the BCN edges.

The amplitudes at the $b$-sublattice of 0th low are given by
\begin{widetext}
\begin{equation}
\bm{\psi}_{b,0} = 
\frac{1}{e^{ik}-1}
\left(
\begin{array}{c}
E_2 \psi_{(a,0,2)} \eta - E_3 \psi_{(a,0,3)} \eta ^3
+ E_4 \psi_{(a,0,4)} \eta^5 - E_1 \psi_{(a,0,1)} \eta^7 \\
E_3 \psi_{(a,0,3)} \eta - E_4 \psi_{(a,0,4)} \eta ^3
+ E_1 \psi_{(a,0,1)} \eta^5 - E_2 \psi_{(a,0,2)} \eta^7 \\
E_4 \psi_{(a,0,4)} \eta - E_1 \psi_{(a,0,1)} \eta ^3
+ E_2 \psi_{(a,0,2)} \eta^5 - E_3 \psi_{(a,0,3)} \eta^7 \\
E_1 \psi_{(a,0,1)} \eta - E_2 \psi_{(a,0,2)} \eta^3
+ E_3 \psi_{(a,0,3)} \eta^5 - E_4 \psi_{(a,0,4)} \eta^7 \\
\end{array}
\right).
\end{equation}
\end{widetext}
For $\bm{\psi}_{a,0}=\bm{C}_{a,1}$, we obtain
\begin{eqnarray}
&&(\bm{C}_{b,2},T_b E_{\rm diag}\bm{C}_{a,1}) \nonumber \\
&=& \frac{(\eta^7+\eta^5+\eta^3+\eta)(E_1-iE_2-E_3+iE_4)}{4(e^{ik}-1)}
\label{eq:L4-a}
\end{eqnarray}
and
\begin{eqnarray}
&& (\bm{C}_{b,3},T_b E_{\rm diag}\bm{C}_{a,1}) \nonumber \\
&=& \frac{(i\eta^7-\eta^5-i\eta^3+\eta)(E_1+E_2+E_3+E_4)}{4(e^{ik}-1)}.
\label{eq:L4-b}
\end{eqnarray}
For $\bm{\psi}_{a,0}=\bm{C}_{a,2}$, we obtain
\begin{eqnarray}
&& (\bm{C}_{b,2},T_b E_{\rm diag}\bm{C}_{a,2}) \nonumber \\
&=& \frac{(\eta^7+\eta^5+\eta^3+\eta)(E_1+E_2+E_3+E_4)}{4(e^{ik}-1)}
\label{eq:L4-c}
\end{eqnarray}
and
\begin{eqnarray}
&&(\bm{C}_{b,2},T_b E_{\rm diag}\bm{C}_{a,2}) \nonumber \\
&=& \frac{(i\eta^7-\eta^5-i\eta^3+\eta)(E_1+iE_2-E_3-iE_4)}{4(e^{ik}-1)}.
\label{eq:L4-d}
\end{eqnarray}
Therefore, these conditions are summarized to two equations. i.e.,
$E_1+E_2+E_3+E_4=0$, and $E_1-E_3=E_2-E_4=0$.
For the B$_\vee$N$_\vee$C$_\vee$C, B$_\vee$C$_\vee$N$_\vee$C and B$_\vee$B$_\vee$N$_\vee$N structures, the former is satisfied, but the latter is not satisfied.
To determine the length of the flat bands, however, we need to care about the wavevector, $k$.

The B$_\vee$N$_\vee$C$_\vee$C, B$_\vee$C$_\vee$N$_\vee$C and B$_\vee$B$_\vee$N$_\vee$N structures do not satisfy Eqs.\  (\ref{eq:L4-a}) and (\ref{eq:L4-b}) because they have the same $k$ region of flat bands.
Therefore, $\bm{\psi}_{a,0}$ is not allowed to have the $\bm{C}_{a,1}$ component.
On the other hand, the $\bm{C}_{a,2}$ component is allowed with $0<k<2\pi/3$ region due to a prohibiting of flat bands in $2\pi/3 <k<\pi$ region by Eq.\ (\ref{eq:L4-d}).
Therefore, the flat bands exist for $0<k<2\pi/3$ for these cases, showing good agreement with the numerical results shown in Fig.\ \ref{fg:L4-band} (b).
The corresponding amplitudes in the $M$th low are given by 
\begin{equation}
\bm{\psi}_{a,m} = \lambda_{a,2}^m \bm{C}_{a,2},
\end{equation}
and
\begin{equation}
\bm{\psi}_{b,m} = \sum_{l=1,3,4}\lambda_{b,l}^m \bm{C}_{a,l}(\bm{C}_{b,l},\bm{\psi}_{b,0}).
\end{equation}

\subsection{$L=6$ case}

In the following, we shall consider the semi-infinite BCN sheet with $L=6$.
As shown in Fig.\ \ref{fg:lambda-2346} (a)-(iv), $|\lambda_{a,2}|$ and $|\lambda_{a,3}|$ are smaller than unity.
Therefore, there are two possibility of the $\bm{\psi}_{a,0}$,
\begin{equation}
\bm{\psi}_{a,0} = \bm{C}_{a,2} \quad\mbox{or}\quad \bm{C}_{a,3}.
\end{equation}
On the other hand, $|\lambda_{b,3}|$ and $|\lambda_{b,4}|$ are larger than unity as shown in Fig.\ \ref{fg:lambda-2346} (b)-(iv).
Therefore, we shall consider the following inner products: $(\bm{C}_{b,3},\bm{\psi}_{b,0})$ and $(\bm{C}_{b,4},\bm{\psi}_{b,0})$ for $\bm{\psi}_{a,0}=\bm{C}_{a,2}$ and $\bm{C}_{a,3}$.

After the direct calculations, we obtain
\begin{widetext}
\begin{eqnarray}
(\bm{C}_{b,3},\bm{\psi}_{b,0})&=& 
\frac{(\eta+\eta^3+\eta^5+\eta^ 7+\eta^9+\eta^{11})
(E_1-\omega E_2+\omega^2 E_3 -E_4+\omega E_5-\omega^2 E_6)}{6(e^{ik}-1)}, \\
(\bm{C}_{b,4},\bm{\psi}_{b,0})&=& 
\frac{(\eta-\omega\eta^3+\omega^2\eta^5-\eta^7+\omega\eta^9-\omega^2\eta^{11})
(E_1+E_2+E_3+E_4+E_5+E_6)}{6(e^{ik}-1)},
\end{eqnarray}
for $\bm{\psi}_{a,0} = \bm{C}_{a,2}$, and 
\begin{eqnarray}
(\bm{C}_{b,3},\bm{\psi}_{b,0})&=& 
\frac{(\eta+\eta^3+\eta^5+\eta^7+\eta^9+\eta^{11})
(E_1+E_2+E_3+E_4+E_5+E_6)}{6(e^{ik}-1)}, \\
(\bm{C}_{b,4},\bm{\psi}_{b,0})&=& 
\frac{(\eta-\omega\eta^3+\omega^2\eta^5-\eta^7+\omega\eta^9-\omega^2\eta^{11})
(E_1-\omega^2 E_2+\omega E_3 -E_4+\omega^2 E_5-\omega E_6)}{6(e^{ik}-1)},
\end{eqnarray}
\end{widetext}
for $\bm{\psi}_{a,0} = \bm{C}_{a,3}$.
Therefore, we obtain the condition for the edge states:
\begin{eqnarray}
\label{eq:L6-1}
&& E_1+E_2+E_3+E_4+E_5+E_6= 0,\\
\label{eq:L6-2}
&& E_1-E_4 = E_3-E_6 = E_5- E_2. 
\end{eqnarray}

The equation (\ref{eq:L6-1}) guarantees the appearance of flat bands if the number of B and N atoms are equal. 
On the other hand, the B$_\vee$N$_\vee$C$_\vee$B$_\vee$N$_\vee$C and B$_\vee$N$_\vee$B$_\vee$N$_\vee$B$_\vee$N structures satisfy the Eq.\ (\ref{eq:L6-2}) equation, while the B$_\vee$N$_\vee$C$_\vee$C$_\vee$C$_\vee$C and B$_\vee$C$_\vee$N$_\vee$C$_\vee$C$_\vee$C structures do not satisfy.

\end{document}